\documentclass[%
 aip,
 amsmath,amssymb,
 reprint,%
]{revtex4-1}

\usepackage{graphicx}% Include figure files
\usepackage{dcolumn}% Align table columns on decimal point
\usepackage{bm}% bold math
    \usepackage{braket}

\usepackage[utf8]{inputenc}
\usepackage[T1]{fontenc}
\usepackage{mathptmx}
\usepackage{etoolbox}
\usepackage{mhchem}

\makeatletter
\def\@email#1#2{%
 \endgroup
 \patchcmd{\titleblock@produce}
  {\frontmatter@RRAPformat}
  {\frontmatter@RRAPformat{\produce@RRAP{*#1\href{mailto:#2}{#2}}}\frontmatter@RRAPformat}
  {}{}
}%
\makeatother
\begin{document}

\preprint{AIP/123-QED}

\title[Seniority-zero  Linear Canonical Transformation Theory]{Seniority-zero Linear Canonical Transformation Theory}
% Force line breaks with \\
\author{Daniel F. Calero-Osorio}
\author{Paul W. Ayers}%
\altaffiliation[Also at ]{ayers@mcmaster.ca}
\affiliation{ 
Department of Chemistry, McMaster University, Hamilton, Ontario L8S 4M1, Canada%\\This line break forced with \textbackslash\textbackslash
}%

\date{\today}% It is always \today, today,

\begin{abstract}
We propose a method to solve the electronic Schrödinger equation for strongly correlated systems by applying a unitary transformation to reduce the complexity of the physical Hamiltonian. In particular, we seek a transformation that maps the Hamiltonian into the seniority-zero space: seniority-zero wavefunctions are computationally simpler, but still capture strong correlation within electron pairs. The unitary rotation is evaluated using the Baker–Campbell–Hausdorff (BCH) expansion, truncated to two-body operators through the operator decomposition strategy of canonical transformation (CT) theory, which rewrites higher-rank terms approximately in terms of one- and two-body operators. Unlike conventional approaches to CT theory, the generator is chosen to minimize the size of non-seniority-zero elements of the transformed Hamiltonian. Numerical tests reveal that this Seniority-zero Linear Canonical Transformation (SZ-LCT) method delivers highly accurate results, usually with submilliHartree error. The effective computational scaling of SZ-LCT is $\mathcal{O}(N^8/n_c)$, where $n_c$ is the number of cores available for the computation.
\end{abstract}

\maketitle

\section{\label{sec:level1} Introduction}

In the orbital picture, spin orbitals are classified as either completely occupied or empty, and the wave function of the physical system is approximated by a single Slater determinant. While this single-determinant description is a useful approximation for weakly correlated systems, it becomes qualitatively inadequate in the strong-correlation regime, where the classification of orbitals as occupied or unoccupied becomes ambiguous. Describing the wavefunctions of such systems, with strong multiconfigurational character, requires summing over multiple Slater determinants.

The multi-configurational nature of strongly correlated systems arises from two fundamental underlying factors. First, due to instantaneous electronic repulsion, electrons can rapidly move between orbitals, causing significant fluctuations in orbital occupancy. The mathematical description of such a wave function is necessarily multi-configurational. This phenomenon is associated with strong dynamic correlation, which can be modeled by Slater determinants representing single, double, triple, and higher-order excitations relative to a chosen reference, typically a single Slater determinant.
Standard approaches for modeling dynamic correlations include configuration interaction (CI),\cite{doi:https://doi.org/10.1002/9781119019572.ch11,szabo1996modernCI,Lowdin1955,KNOWLES1984315,OlsenRoos1988,SherrillSchaefer1999,ShavittBartlett2009,SzalayMuller2012} coupled cluster (CC),\cite{ShavittBartlett2009,doi:https://doi.org/10.1002/9781119019572.ch13,szabo1996modernCC,Coester1960,Cizek1966,paldus1982relationship,CrawfordSchaefer2000,BartlettMusial2007} and many body perturbation theory (MBPT).\cite{ShavittBartlett2009,PhysRev.46.618,doi:https://doi.org/10.1002/9781119019572.ch14,szabo1996modernMBPT,goldstone1957derivation,Bloch1958,FetterWalecka1971,LindgrenMorrison1985} 
The second source of multi-configuration character is near-degeneracies in the system's electronic configurations. In these cases, the multiple Slater determinants do not correspond to excitations from a reference but to a superposition of multiple, nearly isoenergetic system configurations; this type of correlation is usually referred to as static (or nondynamic) correlation. Standard approaches for modeling static correlations include complete active space self-consistent field (CASSCF), multiconfiguration self-consistent field (MCSCF),\cite{RUEDENBERG198241,doi:https://doi.org/10.1002/9780470142943.ch7, 10.1143/PTPS.40.37,Wahl1977} and tensor network state methods (e.g., the density matrix renormalization group (DMRG)).\cite{10.1063/1.478295,10.1063/1.1638734} 
%Add more citacions cor CASSCF, MCSCF and DMRG

\par
A key challenge in contemporary quantum chemistry is the development of methods capable of modeling both dynamic and static correlation within a single framework. To this end, multireference extensions of perturbative approaches originally formulated for single-reference wave functions have been proposed. In particular, complete active space second-order perturbation theory (CASPT2),\cite{Roos1982,doi:10.1021/j100377a012,Andersson1992,Andersson1995,Forsberg1997,Finley1998,Ghigo2004} multireference M\o ller–Plesset theory (MRMP),\cite{HIRAO1992374,Hirao1993,WolinskiPulay,Wolinski,Nakano1993,GrimmeWaletzke2000,witek2002intruder} and $n$-electron valence state perturbation theory (NEVPT2)\cite{10.1063/1.1361246,Angeli2001b,angeli2002n,Angeli2004a} are all based on a multiconfigurational reference (typically CASSCF), which accounts for static correlation, while dynamic correlation is recovered through a second-order perturbative treatment analogous to MP2. However, these multireference perturbation theories often exhibit less favorable computational scaling than their single-reference counterparts, and in the case of CASPT2 and MRMP they can also suffer from intruder-state problems that lead to divergences in the perturbation expansion.\cite{PhysRevA.54.343}

Other extensions of standard single reference methods are multireference configuration interaction (MRCI)\cite{BuenkerPeyerimhoff1974,10.1063/1.455556,10.1063/1.439365} or multireference coupled cluster (MRCC).\cite{JeziorskiMonkhorst1981,RittbyBartlett1991,PaldusPiecuchPylypowJeziorski1993,mahapatra1999size,lyakh2012multireference,kohn2013state} In MRCC, the independent application of the cluster operators to all reference determinants usually create redundancy problems,\cite{ivanov2011state,https://doi.org/10.1002/jcc.540141112} and the algorithms implemented to handle this issue are typically complicated. Moreover, some of these methods, depending on their formulation, can give amplitude equations that are difficult to converge.\cite{ivanov2011state} On the other hand, MRCI is not size-extensive, though this can partially corrected with (generalized) Davidson corrections.\cite{paldusCoupledpairTheoriesDavidsontype1988,
erturkPosterioriCorrectionsConfiguration2015,
meissnerCoupledclusterCorrectionMultireference1999,
szalayConfigurationInteractionCorrections2005} MRCI is usually the method with best accuracy for small systems, however when the number of reference determinants increases it suffers from poor convergence and scalability.\cite{stampfuss1999parallel} \par
The work presented here belongs to a category of methods usually referred as Hamiltonian transformations. The idea behind these methods is that, instead of struggling to model the wavefunction of a strongly-correlated system, one instead aims to transform the system's Hamiltonian so that a less-sophisticated wavefunction ans\"{a}tze will suffice. Examples of such methods include canonical diagonalization (CD; which uses generalized Jacobi rotations to eliminate Hamiltonian couplings whose energy difference exceed a specified cutoff\cite{10.1063/1.1508370} ), the driven similarity renormalization group (DSRG, which employs unitary rotations to suppress selected off-diagonal elements\cite{10.1063/1.4890660,doi:10.1021/acs.jctc.5b00134,10.1063/1.4947218,10.1063/5.0059362}); and canonical transformation (CT) theory (which introduces dynamic correlation on top of a multireference wave function via a tailored unitary transformation\cite{10.1063/1.3086932,10.1063/1.2196410,C2CP23767A,Neuscamman01042010}). \par
Inspired by these methods, this work applies a unitary mapping to recast the molecular Hamiltonian in seniority-zero form. In seniority-zero Hamiltonians, there are no terms that break pairs of electrons, so there are eigenfunctions in which all spatial orbitals are either doubly occupied or empty. By restricting to seniority-zero configurations, the Hilbert-space dimension shrinks to roughly the square root of the full CI space's dimension, greatly simplifying diagonalization.\cite{10.1063/1.3613706,henderson2014seniority,kossoski2022hierarchy,alcobaHybridConfigurationInteraction2014a,CaleroOsorio2025SZ} Beyond this compactness, the seniority-zero Hamiltonian admits a natural mapping onto hard-core bosons/qubits, which is a promising new direction for modelling (strong) electron correlation on quantum computers.\cite{del2025hybrid,krompiec2025simple,costa2025quasiparticle,halderEfficientQuantumState2025,
elfvingSimulatingQuantumChemistry2021a,
leeGeneralizedUnitaryCoupled2019a,
zhaoOrbitaloptimizedPaircorrelatedElectron2023a} Furthermore, the obvious candidate for a reference wave function in the method is a seniority-zero wave function, that as we will explain later in section \ref{sz-reference}, is a special type of multireference wave function for which evaluations of the reduced density matrices (RDMs) and expectation values of operators are especially efficient. Finally, all seniority-zero states can be exactly modelled as a (number-symmetry-broken) geminal mean field,\cite{10.1063/5.0296924} with obvious benefits for interpretability of wavefunction approximations. Indeed, this work is motivated by the recognition that the ground state of seniority-zero systems can be accurately modeled by low-cost geminal mean-field approaches.\cite{doi:10.1021/ct300902c,PhysRevB.89.201106,10.1063/1.4880819,johnsonRichardsonGaudinMeanfield2020,
johnsonSizeconsistentApproachStrongly2013,
tecmerAssessingAccuracyNew2014a,tecmerGeminalbasedElectronicStructure2022,surjanIntroductionTheoryGeminals1999,
surjanStronglyOrthogonalGeminals2012,
zobokiCompositeParticlesQuantum2013b,pernalEquivalencePirisNatural2013b,cullenGeneralizedValenceBond1996,pirisExploringPotentialNatural2024,
pirisPerspectiveNaturalOrbital2014b,johnsonRichardsonGaudinStates2024,
johnsonSingleReferenceTreatment2023,10.1063/5.0088602}
\par

The remainder of this paper is organized as follows. We begin in Sec. \ref{spinfree} with a brief overview of spin-free formalism. Then,  we review the theoretical background of the method in Section \ref{SZ-LCT}. Sec. \ref{sz-reference} explains our choice of the seniority-zero reference wave function, highlighting its lower cost for RDM evaluation. Details on our computational implementation are given in section \ref{implementation}. In Sec. \ref{results} we apply the method to three molecules: \ce{H_6}, \ce{BeH2}, and \ce{BH}. Finally, Sec. \ref{conclusions} offers our conclusions and outlines perspectives for future work.   

\section{Theory}
\label{Theory}

\subsection{Spin-free operators}
\label{spinfree}
Before presenting the method, let us introduce the spin-free formalism, a framework we will use throughout this work.

The creation and annihilation  spin-free operators are defined by tracing over the spin degrees of freedom of the standard spin-orbital creation/annihilation operators:
\begin{equation}
\begin{aligned}
    \hat{E}^{p_1}_{q_1}&=\sum_{\sigma= \alpha, \beta}\hat{c}^{\dagger}_{p_1 \sigma}\hat{c}_{q_1 \sigma}, \\[8pt]
    \hat{E}^{p_1 p_2}_{q_1 q_2}=&\sum_{\sigma,\tau= \alpha, \beta}\hat{c}^{\dagger}_{p_1 \sigma}\hat{c}^{\dagger}_{p_2 \tau}\hat{c}_{q_2 \tau}\hat{c}_{q_1 \sigma},\\[8pt]
    \hat{E}^{p_1 p_2 p_3}_{q_1 q_2 q_3}=&\sum_{\sigma,\tau, \nu= \alpha, \beta}\hat{c}^{\dagger}_{p_1 \sigma}\hat{c}^{\dagger}_{p_2 \tau}\hat{c}^{\dagger}_{p_3 \nu}\hat{c}_{q_3 \nu}\hat{c}_{q_2 \tau}\hat{c}_{q_1 \sigma}.
\end{aligned}
    \label{eq8}
\end{equation}
where $\hat{c}^{\dagger}_{p_1 \sigma}$ ($\hat{c}_{p_1 \sigma}$) represent the creation (annihilation) operator of the spatial orbital $p$ with spin $\sigma$. The reduced density matrices are defined similarly by tracing over the spin degrees of freedom:
\begin{equation}
\begin{aligned}
    \Gamma^{p_1}_{q_1} &=\langle\Psi |\hat{E}^{p_1}_{q_1} |\Psi\rangle,\\[8pt]
    \Gamma^{p_1 p_2}_{q_1 q_2}&=\langle\Psi|  \hat{E}^{p_1 p_2}_{q_1 q_2}|\Psi\rangle,\\[8pt]
    \Gamma^{p_1 p_2 p_3}_{q_1 q_2 q_3}&=\langle\Psi| \hat{E}^{p_1 p_2 p_3}_{q_1 q_2 q_3} |\Psi\rangle.
\end{aligned}
    \label{eq9}
\end{equation}
Since these operators and density matrices don't have spin, their symmetries are different from corresponding operators in the spin-orbital representation. For example:
\begin{equation}
    \begin{aligned}
         E^{p_1 p_2}_{q_1 q_2}= E^{p_2 p_1}_{q_2 q_1}, ~~~~~~~~~~~~~~ E^{p_1 p_2}_{q_1 q_2}= E^{q_1 q_2}_{p_1 p_2}, \\[5pt]
        E^{p_1 p_2}_{q_1 q_2} \neq -E^{p_2 p_1}_{q_2 q_1}, ~~~~~~~~~~~~~~ E^{p_1 p_2}_{q_1 q_2} \neq -E^{p_1 p_2}_{q_2 q_1},
    \end{aligned}
        \label{eq2.17}
\end{equation}
The same (lack of) symmetries hold for the reduced density matrices.

\subsection{The SZ-LCT method}
\label{SZ-LCT}
%talk about the scaling 
We start with the Hamiltonian of a physical system $\hat{H}$, described in second quantization with spin-free operators as:
\begin{equation}
\begin{aligned}
\hat{H}= \sum_{p,q}h_{pq}\hat{E}^{p}_{q} +\frac{1}{2}\sum_{p,q,r,s}v_{pqrs}\hat{E}^{pq}_{rs},
\end{aligned}
    \label{eq1}
\end{equation}
where $h_{pq}$ and $v_{pqrs}$ are the one- and two-body electron integrals respectively, and we used the shorthand notation $\hat{E}^{p_1p_2p_3,...,p_n}_{q_1q_2q_3,...,q_n}=\hat{E}^{\dagger}_{p_1}\hat{E}^{\dagger}_{p_2}\hat{E}^{\dagger}_{p_3}...\hat{E}^{\dagger}_{p_n}\hat{E}_{q_n}\hat{E}_{q_{n-1}}...\hat{E}_{q_1}$, to represent products of creation $\hat{E}^{\dagger}_{p}$ and annihilation $\hat{E}_{q}$ spin-free operators in the spatial orbitals $p$ and $q$, respectively. \par
We aim to map the Hamiltonian into a seniority-zero form, $\hat{H}_{SZ}$, using a unitary transformation:
\begin{equation}
\begin{aligned}
\hat{H}_{SZ}=e^{\hat{A}}\hat{H}e^{-\hat{A}} ,
\end{aligned}
    \label{eq2}
\end{equation}
where $\hat{H}_{SZ}$ has the following pairing structure;
\begin{equation}
\begin{aligned}
\hat{H}_{SZ}=\sum_{p}h_{p}\hat{E}^{p}_{p}  &+ \frac{1}{2}\sum_{p,q}v_{ppqq} \hat{E}^{p\bar{p}}_{q\bar{q}}\\[3pt] & + \frac{1}{4}\sum_{p \neq q}\left(2v_{pqpq} - v_{pqqp}   \right)\hat{n}_{p}\hat{n}_{q},\end{aligned}
    \label{eq3}
\end{equation}
where  $p$, $\bar{p}$ refer to electrons in the same spatial orbital but different spin and $\hat{n}_{p}=\hat{E}^{p}_{p}$ is the number operator. The generator $\hat{A}$ is an anti-hermitian operator made by the combination of excitation and de-excitation operators:
\begin{equation}
\begin{aligned}
\hat{A} =\sum_{p,q}a_{pq}\left(\hat{E}^{p}_{q} - \hat{E}^{q}_{p} \right) +\frac{1}{2}\sum_{p,q,r,s}a_{pqrs}\left(\hat{E}^{pq}_{rs} -\hat{E}^{rs}_{pq} \right),
\end{aligned}
    \label{eq4}
\end{equation}
where $a_{pq}$ and $a_{pqrs}$ are the generators for one- and two-body amplitudes; recall that $a_{pqrs}$ is anti-symmetric with respect to interchange of $p$ and $q$ or $r$ and $s$.\par
The unitary transformation in equation \ref{eq2} is evaluated using the Baker–Campbell–Hausdorff expansion
\begin{equation}
\begin{aligned}
\hat{\widetilde{H}} = \hat{H} + \left[\hat{H},\hat{A}\right] + \frac{1}{2!} \left[\left[\hat{H},\hat{A}\right],\hat{A}\right]  + \frac{1}{3!} \left[\left[\left[\hat{H},\hat{A}\right],\hat{A}\right], \hat{A}\right] +...,
\end{aligned}
    \label{eq5}
\end{equation}
however, given that the generator $\hat{A}$ contains both excitation and de-excitation operators, the expansion  does not truncate at the fourth order as in single-reference CC theories. For practicality, we need to limit the magnitude of the generator $\hat{A}$, so that the expansion can be truncated by $\sim10$-th order. Another difficulty with the evaluation of the BCH expansion is that the number of particle interactions in the Hamiltonian increases which each additional commutator. To address this, we use the strategy first introduced in CT theory, approximating each commutator by at most two-body interactions, so that the terms in the BCH expansion can be evaluated recursively: 
\begin{equation}\label{recursion}
    \begin{split}
    \hat{\widetilde{H}}^{(0)} &= \hat{H} \\
    \hat{\widetilde{H}}^{(n)} &= \tfrac{1}{n}\left[\hat{\widetilde{H}}^{(n-1)},\hat{A}\right]_{1,2},
    \end{split}
\end{equation}
where the subscripts ${1,2}$ indicate that the 3-body operator $\left[\hat{H}^{(n-1)},\hat{A}\right]$ is approximated by a 2-body operator. This approximation, also introduced in CT theory and based on the ideas from the generalized normal order\cite{mukherjee1997normal,kutzelnigg1997normal,kutzelnigg2007generalized}, is founded on the assumption that the reference wave function close enough to the true wave function for the three-body operator to be accurately approximated in terms of one- and two-electron operators and RDMs. The three-body operator decomposition in the spin-free case contains fewer than 90 terms, while in the spin-orbital representation contains $\sim 300$ terms. This motivated us to use the spin-free formulation for our method, as it yields substantial savings in runtime and memory.

The transformed Hamiltonian then has the form,
\begin{equation}
\begin{aligned}
\hat{\widetilde{H}} &= \hat{H} + \left[\hat{H},\hat{A}\right]_{1,2} + \frac{1}{2!} \left[\left[\hat{H},\hat{A}\right]_{1,2},\hat{A}\right]_{1,2}  +.. \\[10pt] 
&=\sum_{n}\hat{\widetilde{H}}^{n} .
\end{aligned}
    \label{eq6}
\end{equation}
The idea behind this approximation is to re-write each high-order excitation operator in terms of lower-order excitations and reduced density matrices using generalized normal order with respect to a (multi)reference wavefunction, $|\Psi_0 \rangle$\cite{mukherjee1997normal,kutzelnigg1997normal,kutzelnigg2007generalized}. This reference is normally an initial guess for the true wave function ($|\Psi \rangle$) of the Hamiltonian that includes the static correlation. Standard choices include CAS, DMRG, etc. 
Instead, we chose the reference to be the ground state of the seniority-zero sector of the target Hamiltonian $\hat{H}$. This choice is motivated by two features of seniority-zero states: (a) the reduced cost of evaluating their RDMs\cite{poelmansVariationalOptimizationSecondOrder2015a} and (b) their ability to capture many types of strong correlation.\cite{CaleroOsorio2025SZ} For example, seniority-zero wavefunctions, such as the number-projected BCS (i.e., AGP) state, \cite{henderson2020correlating, dukelsky2016structure,10.1063/5.0088602} naturally capture the Cooper-pairing physics of conventional (BCS) superconductors. They also provide impressive accuracy for bond-breaking processes.\cite{10.1063/1.3613706,bytautasSeniorityNumberDescription2015,
alcobaHybridConfigurationInteraction2014a,
veillardCompleteMulticonfigurationSelfconsistent1967,
cookDoublyoccupiedOrbitalMCSCF1975,
carboGeneralMulticonfigurationalPaired1977}  
Since the latter is known to be dominated by strong static/non-dynamic correlation, we consider the seniority-zero reference to account for this part of the correlation, while the rest of the correlation will be included using the unitary mapping. 

In summary, the idea of our approach is to find the anti-hermitian generator $\hat{A}$ that maps the molecular Hamiltonian $\hat{H}$ into a seniority zero Hamiltonian $\hat{H}_{SZ}$ (i.e., eliminating the non-seniority-zero Hamiltonian matrix elements that do not appear in equation \ref{eq3}) while keeping the same (low) energy spectrum. This can be expressed as an explicit optimization problem,
\begin{equation}
\begin{aligned}
&\min_{\hat{A}}\left(|| \hat{\widetilde{H}}_{\text{non-}SZ}||\right), \\[10pt]
&\hat{\widetilde{H}}_{\text{non-}SZ}= \hat{\widetilde{H}}-\hat{\widetilde{H}}_{SZ},
\label{eq7}
\end{aligned}
\end{equation}
where $\hat{\widetilde{H}}_{SZ}$ is the seniority-zero sector of the transformed Hamiltonian $\hat{\widetilde{H}}$.

\subsection{Seniority-zero reference}
\label{sz-reference}

% The choice of a seniority-zero wave function as the reference of the method is sustained in multiple factors. First, As previously discussed, this type of multireference wave function encodes information about the static correlation of the physical system. In fact, a particular class of $SZ$ wave functions, the doubly occupied configuration interaction (DOCI) wave function, describes one of the prototypical examples of static correlation, single-bond dissociation, nearly exactly. 
A key advantage of seniority-zero wave functions is the efficiency with which their reduced density matrices (RDMs) can be computed and stored.\cite{poelmansVariationalOptimizationSecondOrder2015a,rubio-garciaVariationalReducedDensity2019,faribaultReducedDensityMatrices2022,
moissetDensityMatricesSeniorityzero2022,
johnsonBivariationalPrincipleAntisymmetrized2022,head-marsdenActiveSpacePairTwoElectron2020,
head-marsdenPair2electronReduced2017,garrosDeterminationReducedDensity2024} 
For a seniority-zero reference wave function $|\Psi_{SZ}\rangle$, the only non-zero elements of the 1RDM are those which preserve the number of pairs in the reference state, so only diagonal elements contribute:
\begin{equation}
\Gamma^{p}_p= \langle\Psi_{SZ}    |\hat{c}^{\dagger}_{p}\hat{c}_{p}|\Psi_{SZ}\rangle.
    \label{Eq10}
\end{equation}
\par
The two-body RDM has two types of non-zero elements. The first type is a pair-excitation from one spatial orbital to another. The second type captures correlations in the occupation of different spatial orbitals. I.e.,
\begin{equation}
\begin{aligned}
    \Gamma^{p\bar{p}}_{q\bar{q}} &= \langle\Psi_{SZ}    |\hat{c}^{\dagger}_{p}\hat{c}^{\dagger}_{\bar{p}}\hat{c}_{q}\hat{c}_{\bar{q}}|\Psi_{SZ}\rangle, \\
    \Gamma^{pq}_{pq} &= \langle\Psi_{SZ}|\hat{c}^{\dagger}_{p}\hat{c}^{\dagger}_{q}\hat{c}_{p}\hat{c}_{q}|\Psi_{SZ}\rangle,
\end{aligned}
    \label{Eq11}
\end{equation}
The elements $\Gamma^{p\bar{p}}_{q\bar{q}}$ are called pair-correlation terms and the elements $\Gamma^{pq}_{pq}$ are called diagonal elements. 
Evaluating a seniority-zero 2RDM ($SZ$-2RDM) has the same computational scaling as evaluating the 1RDM for a generic non-seniority-zero wave function, which is why seniority-zero 2RDM methods have favorable computational scaling.\cite{poelmansVariationalOptimizationSecondOrder2015a,head-marsdenPair2electronReduced2017} 

For general wave functions, the 3RDM is extremely expensive to compute. However, the evaluation of a seniority-zero 3RDM  only considers 3-body excitations that preserve the number of pairs in $|\Psi_{SZ}\rangle$. The non-zero elements of the $SZ$-3RDM are:
\begin{equation}
\begin{aligned}
\Gamma^{pqr}_{pqr} &=  \langle\Psi_{SZ}|\hat{c}^{\dagger}_{p}\hat{c}^{\dagger}_{q}c^{\dagger}_{r}\hat{c}_{p}\hat{c}_{q}\hat{c}_{r}|\Psi_{SZ}\rangle     \\[5pt]
 \Gamma^{pq\bar{q}}_{pq\bar{q}} &=  \langle\Psi_{SZ}|\hat{c}^{\dagger}_{p}\hat{c}^{\dagger}_{q}c^{\dagger}_{\bar{q}}\hat{c}_{p}\hat{c}_{q}\hat{c}_{\bar{q}}|\Psi_{SZ}\rangle      \\[5pt]  
 \Gamma^{pq\bar{q}}_{pr\bar{r}} &= \langle\Psi_{SZ}|\hat{c}^{\dagger}_{p}\hat{c}^{\dagger}_{q}\hat{c}^{\dagger}_{\bar{q}}\hat{c}_{p}\hat{c}_{r}\hat{c}_{\bar{r}}|\Psi_{SZ}\rangle 
 \label{eq12}
\end{aligned}
\end{equation}
All other elements of $\Gamma^{pqr}_{stu}$ are zero. Notice that the elements in the 3RDM with only two indices, i.e. $\Gamma^{pq\bar{q}}_{pq\bar{q}}$ and $\Gamma^{pq\bar{q}}_{pr\bar{r}}$ are identical to terms in the 2RDM, therefore, these elements do not need to be recomputed again. The only new elements in the 3RDM have the form $\Gamma^{pqr}_{pqr}$. A similar process can be used to find the small number of non-zero blocks in the 4-RDM (see the supplementary material for the explicit expressions of the 4RDM non-zero blocks ).

\section{Implementation}
\label{implementation}
Starting from the 1- and 2-electron integrals,\cite{10.1063/5.0006074,kimGBasisPythonLibrary2024,sunLibcintEfficientGeneral2015} we obtained an initial guess for the seniority-zero wavefunction by performing an orbital-optimized doubly-occupied configuration interaction (oo-DOCI) calculation using a development version of \texttt{PyCI}.\cite{10.1063/5.0219010}

To perform the Hamiltonian transformation, we first obtain a symbolic formula for the operator decomposition $\left[\hat{H},\hat{A} \right]_{1,2}$, using an improved version of the \texttt{sqa} software package \cite{10.1063/1.3086932}, which we translated to Python3 and extended to include new features to support the spin-free calculations we are using. Second, we implemented a software package to parse the symbolic expression and evaluate the recursive transformation in eq. \ref{recursion} using \texttt{Numpy} einsum and \texttt{opt\_einsum} to evaluate the tensor contractions\cite{daniel2018opt}. Finally, for the optimization we designed a function that computes the norm of the non-seniority-zero elements of the transformed Hamiltonian; we pass that function to a \texttt{Scipy} minimizer. As was mentioned before, in order to make the transformation of Eq. \ref{eq6} accurate, a small generator $\hat{A}$ needs to be used. For that reason we conduct the minimization with a constraint over the norm of the generator ($||\hat{A}|| \leq \epsilon$), where this epsilon is determined leveraging previous knowledge of the DOCI prediction compared to the exact energy. We used \textit{SLSQP} and \textit{trust-constr} algorithms for the minimization, as they allow non-linear constraints. At the end of the optimization, the minimizer returns the generator $\hat{A}$ that minimizes the norm of the non-seniority-zero elements of the Hamiltonian. As part of this work, we extended \texttt{PyCI} to efficiently evaluate RDMs for seniority-zero wavefunctions; these RDMs are then used when evaluating the decomposition $\left[\hat{H},\hat{A} \right]_{1,2}$.

\subsection{Cost and performance}
As mentioned in the previous section, our approach requires an orbital-optimized DOCI (oo-DOCI) calculation to provide the seniority-zero reference. The cost of this step scales as the number of optimization iterations multiplied by the square root of the cost of a full configuration interaction (FCI) calculation in the same orbital basis. Although this can become a burden for large systems, several approximate oo-DOCI schemes with polynomial or even mean-field-like scaling have been proposed in the literature \cite{surjanIntroductionTheoryGeminals1999,rassolov2002geminal,surjan2012strongly,parr1956generalized,parks1958theory,kutzelnigg1964direct,jeszenszki2015local,pernal2013equivalence,hurley1953molecular, parks1958theory, hunt1972self, hay1972generalized, small2014coupled,lawler2008symmetry,cullen1996generalized,moss1975generalized,dykstra1980perfect,carter1988correlation,hartke1992ab,doi:10.1021/ct300902c,PhysRevB.89.201106,10.1063/1.4880819,Boguslawski2015,JOHNSON2013101,moissetDensityMatricesSeniorityzero2022,faribaultReducedDensityMatrices2022,fecteauNearexactTreatmentSeniorityzero2022,fecteauReducedDensityMatrices2020,johnsonBivariationalPrincipleAntisymmetrized2022,johnsonRichardsonGaudinMeanfield2020,johnsonRichardsonGaudinStates2024,tecmerAssessingAccuracyNew2014a,henderson2025jordanwignertransformationdescriptionstrong} , and the formalism developed here can be straightforwardly combined with such seniority-zero reference wavefunctions.

Now, to assess the computational cost of the method, we focus on its most expensive components. The first is the evaluation of the decomposition $\left[\hat{H},\hat{A} \right]_{1,2}$,  which in general scales as $\mathcal{O}(N^7)$, with $N$ the number of orbitals. The second is the gradient evaluation during the optimization, which typically scales as $\mathcal{O}(N^4)$. Direct implementation of this procedure, then, would scale as $\mathcal{O}(N^{11})$. To reduce this cost, two major improvements were implemented. First, we exploit the structure of the seniority-zero RDMs by evaluating the tensor contractions in $\left[\hat{H},\hat{A} \right]_{1,2}$ using only the non-zero elements, rather than the full RDMs. This reduces the scaling of the operator decomposition from $\mathcal{O}(N^7)$ to $\mathcal{O}(N^5)$. Second, for the gradient, we developed an analytical implementation and parallelized its evaluation so that $n_c$ gradient components can be computed concurrently, where $n_c$ is the number of available cores. Moreover, due to the antisymmetry of the two-body amplitudes of the generator $A$, the total number of independent elements scales as $\mathcal{O}\left(\frac{N^2 (N-1)^2}{4}\right)$. For fewer than approximately 150 spatial orbitals this is effectively $\mathcal{O}(N^3)$ scaling, so the gradient cost becomes $\mathcal{O}\left(\frac{N^3}{n_c}\right)$ for small- and moderate-sized systems. With these optimizations, the effective overall scaling of the method is reduced to $\mathcal{O}\left(\frac{N^8}{n_c}\right)$. 

The closest CT analog to our method is the linear canonical transformation theory with singles and doubles, including explicitly the 3-RDM (L3CSTDS)\cite{10.1063/1.3086932,Neuscamman01042010}, which scales as the seventh power of the number of active-space orbitals $n_a$. Although the scaling of this method is favorable, SZ-LCT has some advantages over L3CSTDS. First, as discussed in the previous section, the spin-free representation significantly reduces the scaling prefactor compared to the spin-orbital representation of L3CSTDS. Furthermore, while the parallelization in SZ-LCT is focused on reducing the computational scaling, in L3CSTDS the parallel algorithm is designed for distributing the memory associated with tensors up to size $n_a^6$, which is not necessary in our case since we only store the non-zero elements of the seniority-zero RDMs.

\section{Results}
\label{results}
\subsection{\ce{H_6}} 
\begin{figure}[h]
\includegraphics[width=0.97\linewidth]{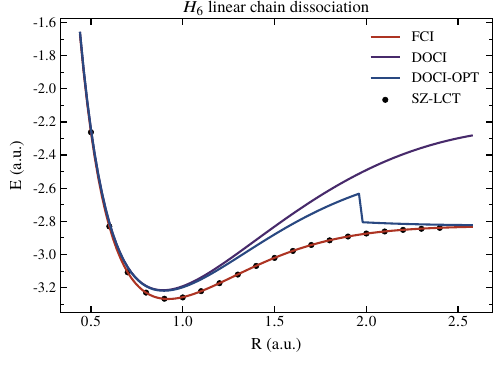} 
\caption{\label{fig1} Dissociation curve for the linear \ce{H6} chain in the STO-6G basis set as a function of nearest-neighbor distance.}
\end{figure}

\begin{figure}[h]
\includegraphics[width=0.97\linewidth]{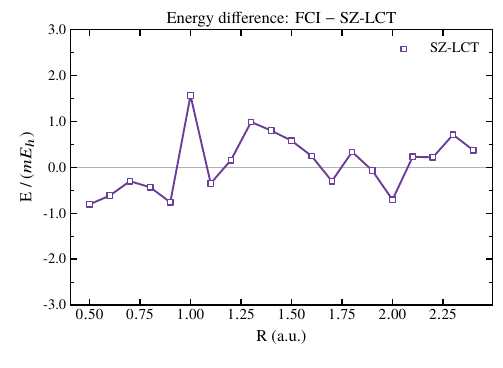} 
\caption{\label{fig2}  Energy difference between the SZ-LCT and FCI, in $mE_h$, for \ce{H6} chain dissociation in the STO-6G basis set}
\end{figure}

As a first test case we stretch the linear \ce{H6} chain in a minimal (STO-6G) basis. Results for the ground state energy dissociation with  SZ-LCT method along with the Full-CI (FCI) and DOCI energies are plotted in figure \ref{fig1}; the energy difference is plotted in figure \ref{fig2}. The reference wave function used for the SZ-LCT method was the orbital-optimized(oo) DOCI (blue curve). While orbital optimization was not essential for near-equilibrium bond lengths, orbital optimization becomes important near dissociation. Therefore, while we had hoped that including one-body operators in SZ-LCT could replace the problematic orbital optimization that afflicts all seniority-zero methods,\cite{boguslawskiProjectedSenioritytwoOrbital2014b,
limacherInfluenceOrbitalRotation2014,boguslawskiNonvariationalOrbitalOptimization2014,szabadosOrbitalOptimisationSpinunrestricted,kossoskiExcitedStatesStateSpecific2021,
fecteauNearexactTreatmentSeniorityzero2022,
kossoski2022hierarchy} we conclude that orbital optimization is still necessary in general. 
\par

For the \ce{H6} symmetric stretch, SZ-LCT performs well, with all errors within chemical accuracy ($<1$kcal/mol $\approx 1.6 mE_h$. Recalling that the energies obtained from the SZ-LCT method correspond to the solutions of the eigenvalue problem $\hat{H}_{SZ}\ket{\Psi_{SZ}}=E_{SZ}\ket{\Psi_{SZ}}$ where $\hat{H}_{SZ}$ is obtained from the unitary transformation \ref{eq2}, this means that we were able to find a seniority-zero Hamiltonian whose seniority-zero ground-state wave function captures the physical behavior of the exact ground-state wave function. 
Notably, the SZ-LCT method performs equally well along the entire potential energy curve; this is especially reassuring since the underlying reference wavefunction gives more accurate energies for compressed and near-dissociation geometries. It is also remarkable that in the vicinity of $R = 1.9 \text{ a.u.}$, where the oo-DOCI computation falls into a local minimum, the SZ-LCT solution remains excellent. The ability of SZ-LCT to produce quantitatively correct results even when oo-DOCI gets trapped in a local minimum is a promising signal for the robustness of this approach.

\par
From figure Fig. 2 we note, contrary to our expectations, the energy error in the method is not continuous. Given that the jumps in the energy difference are, with just one exception, always less than a milliHartree, these jumps might be induced by numerical instabilities. However, upon close inspection, we noticed that the minimizing generator matrices of nearby geometries are dissimilar, indicating that the method is erratically shifting between different, nearly degenerate, local minima. This is not surprising because multireference coupled cluster methods are known to be ill-conditioned and we did not do any single value decomposition to remove near singularities. Further improvements of the method will focus on this direction.

\subsection{\ce{BH}} 
\begin{figure}
\includegraphics[width=0.97\linewidth]{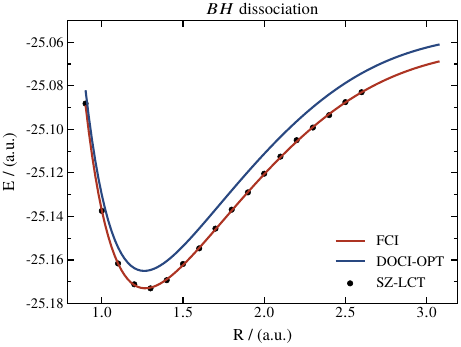} 
\caption{\label{fig3} \ce{BH} dissociation energy curve in 6-31G basis set.}
\end{figure}

\begin{figure}
\includegraphics[width=0.97\linewidth]{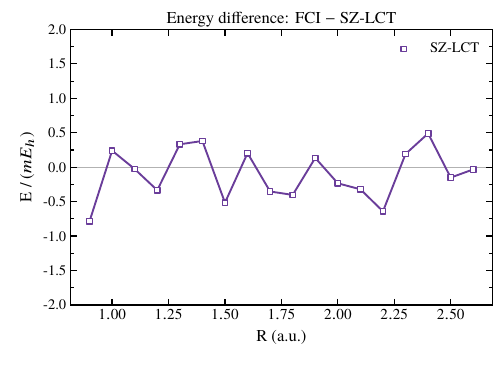} 
\caption{\label{fig4}  Energy difference between the SZ-LCT and FCI, in $mE_h$, for \ce{BH} dissociation.}
\end{figure}

For the second test we chose \ce{BH} dissociation in the 6-31G basis set. As a diatomic single bond molecule with small inter-pair interactions, Boron hydride is very well described by oo-DOCI, as can be seen from figure \ref{fig3}; the maximum energy error in oo-DOCI is $\sim 9mE_h$, which occurs near the equilibrium bond length. Results for the SZ-LCT energy prediction along with FCI results are presented in Figure \ref{fig3}; errors with respect to FCI are plotted in Figure \ref{fig4}. With all errors substantially below $1mE_h$, \ce{BH} shows the best performance among all the studied cases, as might expected given that excellent performance of the reference oo-DOCI calculation. With the 6-31G basis set, \ce{BH} is described by eleven basis functions, therefore, the cost of the method using  $ \sim 100$ cores is comparable to a single reference method. For larger basis sets, when the number of basis functions is larger than the number of available cores $n_c$, our algorithm will require further optimization.

\subsection{\ce{N2}} 
\begin{figure}[h!]
\includegraphics[width=1.0\linewidth]{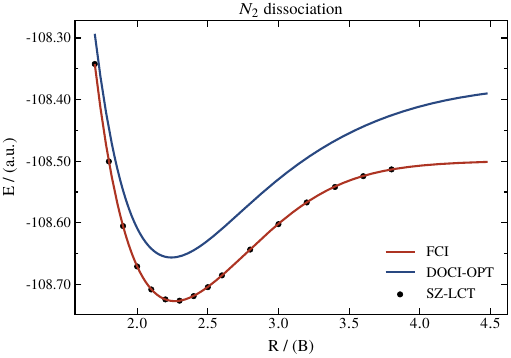} 
\caption{\label{fig5} \ce{N2} dissociation energy curve in STO-6G basis set.}
\end{figure}

\begin{figure}[h!]
\includegraphics[width=1.0\linewidth]{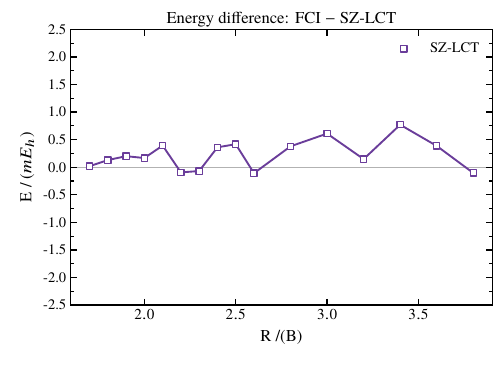} 
\caption{\label{fig6}  Energy difference between the LT-SZCT method and FCI in $mE_h$ for \ce{N2} dissociation in STO-6G basis set.}
\end{figure}

We close this section by presenting \ce{N2} dissociation in the STO-6G basis set. Unlike single-bond dissociation, breaking multiple bonds is a known weakness of DOCI.\cite{bytautasSeniorityNumberDescription2015}. When breaking the triple bond in \ce{N2}, oo-DOCI energy predictions have a stable error of about $8 \times10^{-2} ~ E_h$ for compressed and equilibrium geometries, reaching $\sim 10^{-1}~E_h$ in the dissociation limit (cf. figure \ref{fig5}). Results for the ground state energy from SZ-LCT along with the FCI and oo-DOCI energies are plotted in figure \ref{fig5}; the energy difference between FCI and SZ-LCT is plotted in figure \ref{fig6}.

As shown in Fig.~\ref{fig5},SZ-LCT delivers highly accurate results along the entire dissociation curve, including regimes where oo-DOCI errors reach $\sim 10^{-1}~E_h$. For compressed and near-equilibrium geometries ($R \le 2.5$~\AA), SZ-LCT deviations with respect to FCI remain below $0.5$~m$E_h$. Upon stretching the bond—where oo-DOCI is no longer \textit{quantitatively} reliable—SZ-LCT errors increase slightly but stay below $0.8$~m$E_h$. This highlights the robustness of the approach even in a regime where DOCI is known to be a less reliable reference.\cite{bytautasSeniorityNumberDescription2015} Finally, we note that most SZ-LCT energies lie below the FCI curve, indicating that the transformation is not strictly unitary in due to the recursive commutator approximation used to evaluate Eq.~\ref{eq7}. While this does not compromise the overall accuracy here, as mentioned before, a minor modification of the generator constraint could mitigate this behavior and enforce SZ-LCT energies to remain variational (i.e., above FCI).

\section{Conclusions}
\label{conclusions}
We have presented a method to map general electronic Hamiltonians to simpler, seniority-zero, Hamiltonians. Using seniority-zero wave functions for the reference and efficient parallel algorithms, we reduced the computational cost enough so that this approach can be used for small- to medium-sized molecules when the number of available cores for the computation is no less than the number of spatial orbitals.

The method shows highly accurate energy predictions for the three molecules tested, with results within chemical accuracy. It is important to use orbital-optimized doubly-occupied configuration interaction (oo-DOCI) as a reference: tests with non–orbital optimized DOCI had significantly higher errors, revealing that SZ-LCT requires a reference state that captures key qualitative features of the true wave function. Nevertheless, the method worked well even where oo-DOCI was qualitatively inaccurate (cf. Figure 3) or optimized to a local, rather than the global, minimum (cf. Figure 1).

We hoped that we might mitigate SZ-LCT's need for a qualitatively correct reference wave function by iteratively refining the reference. In this way, instead of attempting to render the Hamiltonian seniority-zero with one transformation as in Equation \ref{eq2}, we drive the Hamiltonian progressively closer to seniority-zero using several transformations:
\begin{equation}
\begin{aligned}
H_1 &=e^{A_1}He^{-A_1}, \\[5pt]
H_2 &=e^{A_2}He^{-A_2}, \\[5pt]
...\\
H_{SZ}&=e^{A_n}He^{-A_n}, \\[5pt]
\end{aligned}
\end{equation}
We attempted to update the reference at each step to the ground-state wave function of the seniority-zero sector of that step’s Hamiltonian, thereby improving the reference quality. (I.e., the second iteration uses the seniority-zero eigenvector of the SZ-LCT Hamiltonian, $\hat{H}_{SZ}$, as its reference wavefunction.) However, our tests showed that applying the commutator approximation repeatedly at each step leads to an accumulation of error, substantially degrading the accuracy of the method. 

Further refinements of this method are certainly warranted. In particular, we have established that the spikes in the FCI-SZ-LCT energy-difference curves are not caused by numerical noise, but by the amplitude solver converging erratically to different, nearly degenerate, local minima of the objective function. A natural direction for future work is to incorporate a singular-value-decomposition (SVD)-based regularization of the excitation manifold in order to stabilize the optimization along the potential energy surface.
Second, we also are looking to further improve the computational efficiency, focusing on the evaluation of the gradient (e.g., by cleverly selecting a subset of generator parameters to update in each gradient evaluation, we might reduce the computational cost significantly.) Finally, we should investigate cases where the seniority-zero Hamiltonian may be a poor choice. 

Ultimately, this method still requires accurate solutions to the seniority-zero problem. Our belief is that recent low-scaling geminal-based approximations will suffice for the seniority-zero problem, but that this Hamiltonian-transformation strategy is a more mathematically elegant and accurate way to add dynamic correlation than previous approaches based on coupled-cluster, density-functional, or perturbative methods.\cite{Boguslawski2015,kobayashiGeneralizedMollerPlessetPartitioning2010,
zobokiLinearizedCoupledCluster2013,margocsyMultipleBondBreaking2018,
pastorczakERPAAPSGComputationallyEfficient2015,
pernalIntergeminalCorrectionAntisymmetrized2014,garzaRangeSeparatedHybrids2015,
garzaSynergyPairCoupled2015,gaikwadCoupledClusterinspiredGeminal2024,
miranda-quintanaFlexibleAnsatzNBody2024,nowakConfigurationInteractionCorrection2023,vuSizeextensiveSeniorityzeroEnergy2020,nowakOrbitalEntanglementCorrelation2021,pirisAssessmentSecondorderPerturbative2014b,
pirisInteractingPairsNatural2014b,vanraemdonckPolynomialScalingApproximations2015a,bulikCanSingleReferenceCoupled2015b} 

\section{Supplementary Material}
Expressions for the nonvanishing matrix elements of the 4-electron reduced density matrix for a seniority-zero state and tabulations of the numerical results are provided as supplementary material.

\begin{acknowledgments}
The authors thank the Canada Research Chairs (CRC-2022-00196), the Digital Research Alliance of Canada, and NSERC (ALLRP/592521-2023 and RGPIN-2024-06707) for financial and computational support.
\end{acknowledgments}

\section*{Data Availability Statement}

The data that support the findings of this study are available from the corresponding author upon reasonable request. Tabulated data supporting the figures is included as supplementary material.

%\appendix

%\section{Appendixes}

%\nocite{*}
\bibliography{aipsamp_clean}% Produces the bibliography via BibTeX.

\end{document}